\documentclass[rnote, traditabstract]{aa}  
\usepackage[english]{babel}
\usepackage{txfonts, graphicx}
\usepackage[usenames]{color}
\usepackage{natbib}
\bibpunct{(}{)}{;}{a}{,}{,}

\begin{document}

\title{A background galaxy in the field of the \object{$\beta$ Pic} debris disk\fnmsep\thanks{\textit{Herschel} is an ESA space observatory with science instruments provided by European-led Principal Investigator consortia and with important participation from NASA.}}
\author{S.~Regibo\inst{\ref{IvS}}\and B.~Vandenbussche\inst{\ref{IvS}}\and
C.~Waelkens\inst{\ref{IvS}}\and  {B.~Acke\inst{\ref{IvS}}\fnmsep\thanks{Postdoctoral Fellow of the Fund for Scientific Research, Flanders}}\and B.~Sibthorpe\inst{\ref{RO Edinburgh}} \and M.~Nottebaere\inst{\ref{IvS}}\and K.~Voet\inst{\ref{IvS}}\and \\ 
J.~Di Francesco\inst{\ref{Canada}}\and M.~Fridlund\inst{\ref{ESTEC}}\and W.K.~Gear\inst{\ref{Cardiff}}\and R.J.~Ivison\inst{\ref{RO Edinburgh}, \ref{IfA Edinburgh}}\and G.~Olofsson\inst{\ref{Stockholm}}}

\institute{Instituut voor Sterrenkunde, KU Leuven, Celestijnenlaan 200D, B-3001 Leuven, Belgium
\newline \email{sara@ster.kuleuven.be}\label{IvS}
\and UK Astronomy Technology Centre, Royal Observatory Edinburgh, Blackford Hill, Edinburgh EH9 3HJ, UK\label{RO Edinburgh}
\and National Research Council of Canada, Herzberg Institute of Astrophysics, 5071 West Saanich Road, Victoria, BC, V9E 2E7, Canada\label{Canada}
\and ESA Research and Science Support Department, ESTEC/SRE-S, Keplerlaan 1, 2201 AZ, Noordwijk, The Netherlands\label{ESTEC}
\and School of Physics and Astronomy, Cardiff University, Queens Buildings The Parade, Cardiff CF24 3AA, UK\label{Cardiff}
\and Institute for Astronomy, University of Edinburgh, Blackford Hill, Edinburgh EH9 3HJ, UK\label{IfA Edinburgh}
\and Department of Astronomy, Stockholm University, AlbaNova University Center, Roslagstullbacken 21, 10691 Stockholm, Sweden\label{Stockholm}
}


\date{Received [date] / Accepted [date]}



\abstract{\textit{Herschel} images in six photometric bands show the thermal emission of the debris disk surrounding \object{$\beta$ Pic}.  In the three PACS bands at 70\,$\mu$m, 100\,$\mu$m and 160\,$\mu$m and in the 250\,$\mu$m SPIRE band, the disk is well-resolved, and additional photometry is available in the SPIRE bands at 350\,$\mu$m and 500\,$\mu$m, where the disk is only marginally resolved.  The SPIRE maps reveal a blob to the southwest of\object{$\beta$ Pic}, coinciding with submillimetre detection of excess emission in the disk.  We investigated the nature of this blob.  Our comparison of the colours, spectral energy distribution and size of the blob, the disk and the background sources shows that the blob is most likely a background source with a redshift between $z =1.0$ and $z = 1.6$.}

\maketitle

\section{Introduction}
The circumstellar disk around the A6V star \object{$\beta$ Pic} was discovered by IRAS in 1983 \citep{1984ApJ...278L..23A} and especially the inner part has been studied extensively ever since \citep{1995AJ....110..794K, 1997A&A...327.1123P, 1997MNRAS.292..896M, 2000ApJ...539..435H, 2005NatureTelesco}. Optical observations imply that the disk extends to $95 \arcsec$ offset, which corresponds to 1800\,AU \citep{2001MNRAS.323..402L}.  Warps and asymmetries in the inner part of the disk suggest the presence of planetesimals and even planets, the latter of which has been confirmed by \citet{2010ScienceLagrange}.

Owing to its relative proximity to the Earth 
\citep[$19.44 \pm 0.05\,\mathrm{pc}$;][]{2007A&A...474..653V}, the debris disk was resolved spatially at longer wavelengths as well \citep{1998Natur.392..788H, 2009A&A...508.1057N, 2003A&A...402..183L}.

The \object{$\beta$ Pic} debris disk has been imaged by \textit{Herschel} in six photometric bands.  In the 250-500\,$\mu$m bands a blob appears to the southwest of the star.  This blob coincides with the one observed at 850\,$\mu$m with SCUBA by \citet{1998Natur.392..788H} and 870\,$\mu$m with LABOCA by \citet{2009A&A...508.1057N} and both teams of authors assumed it to be a structure in the disk.

A first counter-argument was already mentioned in \citet{2000MNRAS.314..702D}.  Assuming the blob indeed originates in the debris disk and has the same dust characteristics, its observed emission at 850\,$\mu$m can only be explained by at least $1.2\,M_{Moon}$, a mass comparable to the total mass of the disk in small dust particles.  Given the blob's distance from the central star and the fact that there is no counterpart for such a large amount of dust in deep optical observations, \citet{2000MNRAS.314..702D} argued that the blob cannot be a structure in the disk. \citet{2010A&A...518L.133V} argued that given the high amount of background galaxies in the neighbourhood of \object{$\beta$ Pic}, the blob probably is a background source.

In this paper, we investigate the hypothesis that the blob is a background source and not a structure in the circumstellar disk.

\begin{figure*}
	\centering
	\includegraphics[width=17cm]{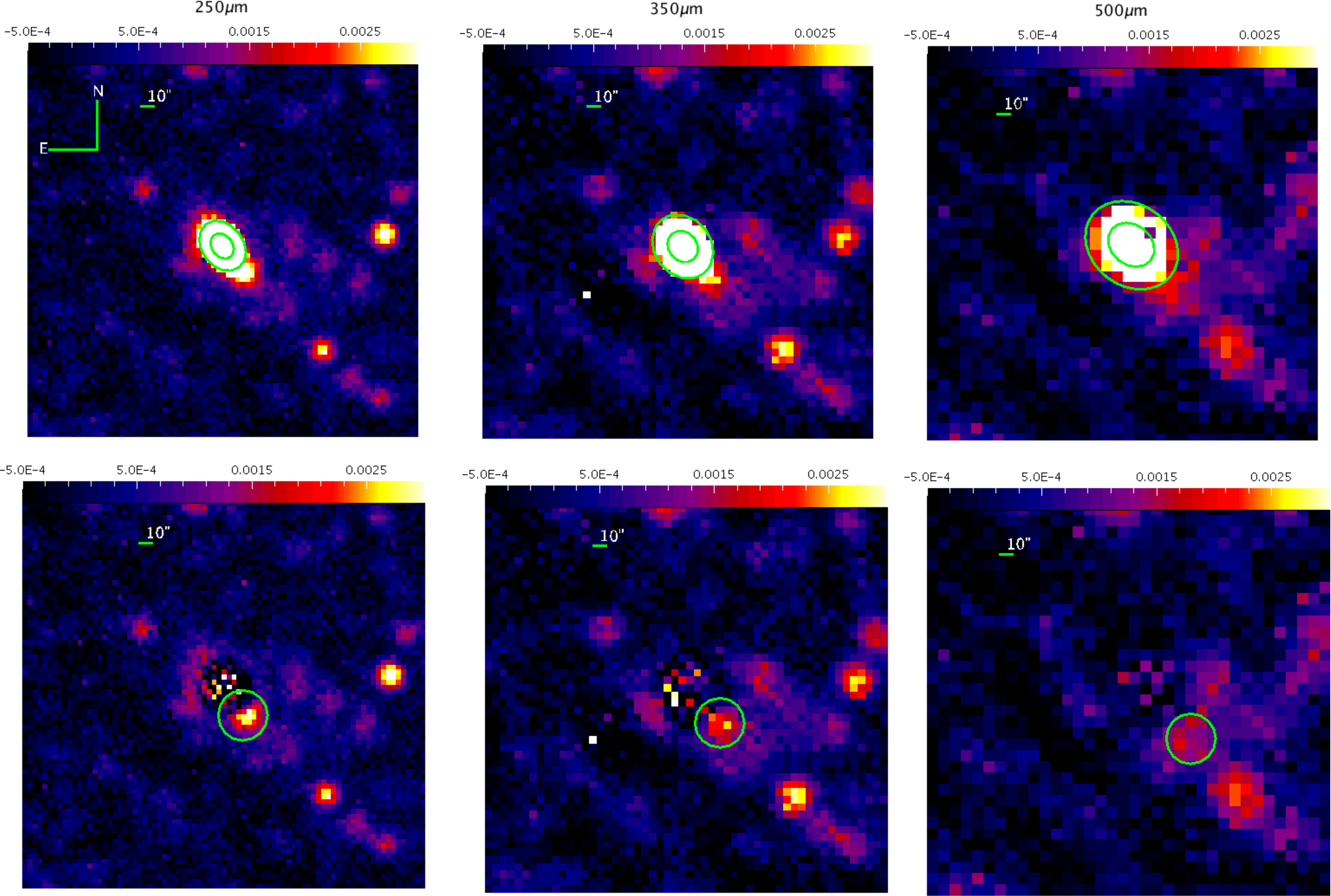}
	\caption{Surface brightness maps in the SPIRE bands (in the upper panels) with the fitted Gaussian overplotted; and the maps after subtraction of the fitted disk, with the position of the blob indicated with a $20 \arcsec$ radius circle in the lower panels (left: 250\,$\mu$m; middle: 350\,$\mu$m; right: 500\,$\mu$m).  All maps display the same area on the sky and are centred on the optical position of \object{$\beta$ Pic}.}
	\label{fig : overview}
\end{figure*}

\section{Observations and data reduction}
Photometric maps were obtained with the \textit{Herschel Space Observatory} \citep{2010A&A...518L...1P} with the PACS and SPIRE photometers.  A concise observing log can be found in Table \ref{table : observations}.  More information concerning the scientific capabilities, obser\-ving modes, data reduction, calibration, and performance of these instruments can be found in \citet{2010A&A...518L...2P} for PACS, and in \citet{2010A&A...518L...3G} and \citet{2010A&A...518L...4S} for SPIRE.
 
The observations were performed in the context of the `Stellar Disk Evolution' Guaranteed Time Key Programme (PI: G.~Olofsson) during the  science demonstration phase.

\begin{table}
	\caption{Observation log for the photometric \textit{Herschel} observations of \object{$\beta$ Pic}. Column 1 gives the instrument, column 2 the observation identification, column 3 the date of the observation, column 4 its duration and column 5 the observed wavelengths.}
	\label{table : observations}
	\centering
	\begin{tabular}{ccccc}
		\hline \hline
		& Observation & Date & Duration & Filters [$\mu$m] \\
		\hline
		SPIRE & 1342187327 & 2009-11-30 & 3336\,s & 250, 350, 500 \\
		PACS & 1342185457 & 2009-10-07 & 808\,s & 100, 160 \\
		PACS & 1342186612 & 2009-11-01 &  5506\,s & 70, 160 \\
		PACS & 1342186613 & 2009-11-01 &  5506\,s & 70, 160 \\
		\hline
	\end{tabular}
\end{table}
 
Pipeline processing as well as image analysis have entirely been made in the interactive processing environment HIPE, which has been developed for the \textit{Herschel} project \citep{2010AAS...21641310O}.

\subsection{SPIRE observations}
The SPIRE observation includes three wavelength bands simultaneously (at 250\,$\mu$m, 350\,$\mu$m and 500\,$\mu$m), covering an area of $8 \arcmin \times 8 \arcmin$ homogeneously. The maps were constructed with the \textit{naiveMapper} task.  This algorithm projects each sample onto a single map pixel.  The pixel size is $4 \arcsec$, $6 \arcsec$ and $9 \arcsec$ in the three bands respectively, and the beam FWHM is $18.2 \arcsec$, $24.9 \arcsec$ and $36.3 \arcsec$ respectively.

\subsection{PACS observations}
At 100\,$\mu$m, we have a single shallow PACS scan map, taken at medium speed ($20 \arcsec \mathrm{s}^{-1}$); while at 70\,$\mu$m and 160\,$\mu$m, a scan and a cross-scan at low speed ($10 \arcsec \mathrm{s}^{-1}$) are at our disposal.  For the shallow observation, an area of $2 \arcmin \times 2 \arcmin$ is covered homogeneously; for the deep observations an area of $2.5 \arcmin \times 2.5 \arcmin$.

The PACS observations were processed with the HIPE pipeline script for the reduction of observations of extended sources, such as \object{$\beta$ Pic}.  Removal of cosmic rays was performed with the \textit{IIndLevelDeglitch} task.  For each pixel, $\sigma$-clipping was applied on all contributions before projecting the remaining values onto a map.  The flux conversion was made with version 6 of the response calibration.  Before applying a high-pass filter to eliminate low-frequency drifts, the region around the source was masked.  The width of the filter window was taken as half the scan leg length ($3.7 \arcsec$).  We used the \textit{photProject} task to combine the detector time series signals into maps of $1\arcsec$ pixels in the blue and green, and $2 \arcsec$ pixels in the red, weighting the signal by the respective noise.

\section{Image analysis}\label{section : image analysis}
In this section we discuss how we determined the flux of the debris disk, the blob, and the background sources.

\begin{figure}
	\resizebox{\hsize}{!}{\includegraphics{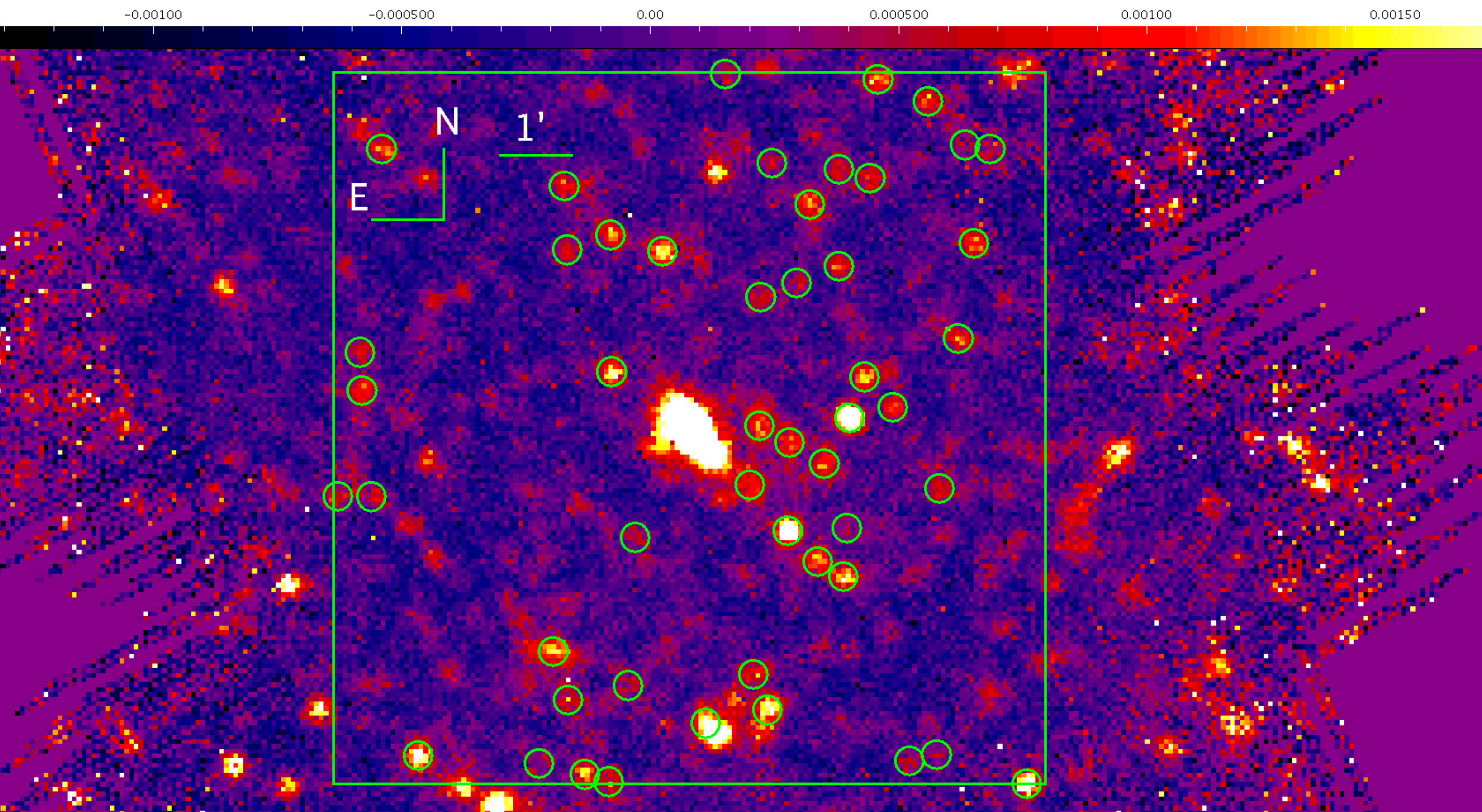}}
	\caption{SPIRE map at 250\,$\mu$m on which the background sources are indicated with green circles.  The green rectangle indicates the region that is covered by the two scan directions.}
	\label{fig : background sources}
\end{figure}

From the SPIRE maps shown in Fig.\,\ref{fig : overview}, a blob to the southwest of the debris disk of \object{$\beta$ Pic} can be clearly seen (at an offset $(\Delta \alpha, \Delta \delta) = (-25 \arcsec,-23 \arcsec)$ from the star), after subtraction of the flux of the disk.  In Fig.\,\ref{fig : background sources}, we notice tens of background sources in the $10 \arcmin \times 10 \arcmin$ region around the star.  We quantified whether the far-IR colours of the blob resemble the colours of these other sources more than the colours of the debris disk itself.  If so, this would be a strong indication that the blob is a background source and not a structure in the disk.

In the PACS maps, the background sources are less prominent, and the complex shape of the PACS point-spread-function (PSF) prohibits a swift separation of the disk flux from the PACS maps.  Moreover, the surface brightness profiles for PACS in \citet[][see Fig.\,3]{2010A&A...518L.133V} suggest the blob is faint and cannot be discerned in the PACS wavelength range.

In the SPIRE maps, we separated the fluxes of the blob from the fluxes of the debris disk by fitting an elongated Gaussian to the observed disk and subtracting it, using the \textit{sourceFitting} and \textit{imageSubtract} tasks in HIPE.  Fig.\,\ref{fig : overview} shows the original SPIRE maps on which the respective fitted Gaussians were overplotted, and the residual maps with the blob indicated.  The fit parameters are summarised in Table \ref{table : fit parameters}.

\begin{figure*}
	\centering
	\includegraphics[width=17cm]{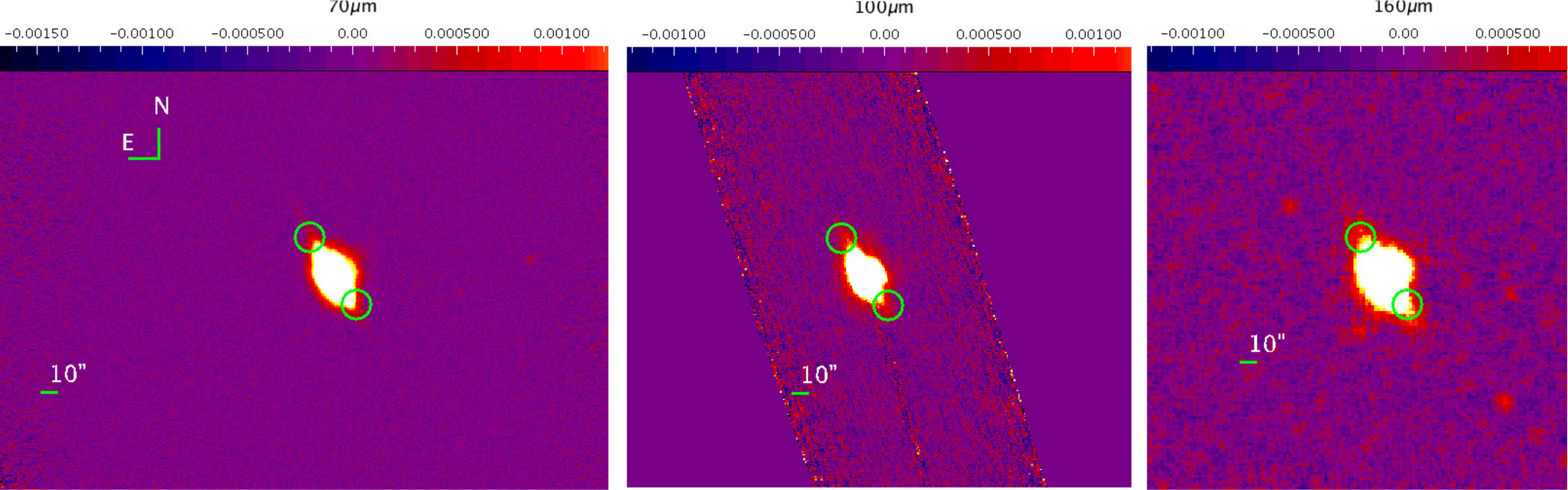}
	\caption{Surface brightness maps in the PACS bands with the $10 \arcsec$ apertures used to determine an upper limit for the flux of the blob overplotted (left : 70\,$\mu$m; middle : 100\,$\mu$m; right : 160\,$\mu$m).}
	\label{fig : PACS maps}
\end{figure*}

\begin{table}
	\caption{Fit parameters for the subtracted Gaussian disk model in the SPIRE wavelength bands.}
	\label{table : fit parameters}
	\centering
	\begin{tabular}{cccc}
		\hline \hline
		Parameter & 250\,$\mu$m & 350\,$\mu$m & 500\,$\mu$m \\
		\hline
		$\alpha_{center}$ [h m s] & $05 \ 47 \ 17.12$ & $05 \ 47 \ 17.11$ & $05 \ 47 \ 16.94$ \\
		$\delta_{center} \ [\degr \ \arcmin \ \arcsec]$& $-51 \ 03 \ 55.5$ & $-51 \ 03 \ 55.52$ & $-51 \ 03 \ 58.20$ \\
		FWHM$_{major \ axis} [\arcsec]$ &  23.20 & 28.96 & 43.01 \\ 
		FWHM$_{minor \ axis} [\arcsec]$ & 17.76 & 23.54  & 34.78\\ 
		Position angle $[\degr]$ & 37.40 & 39.31 & 39.31 \\
		\hline
	\end{tabular}
\end{table}

We can now determine the flux of the debris disk and of the blob by performing aperture photometry on the original maps and on the difference maps with the \textit{rectangularSkyAperturePhotometry} task in HIPE.  This task calculates the flux within a circular aperture and estimates a sky contribution from a rectangular region close enough to the target to achieve a similar coverage.

To calculate the flux of the disk, we used a circular $60 \arcsec$ radius target aperture.  A value for the sky contribution was obtained with a sky estimation algorithm, corresponding to the \textit{mmm} routine from the IDL Astronomy Library to estimate the background in a stellar contaminated field.  The acquired fluxes are summarised in Table\,\ref{table : fluxes disk}.  The errors on the fluxes are do\-minated by uncertainties in flux calibration, which are less than 15\% for SPIRE \citep{2010A&A...518L...4S}, less than 10\% at 70\,$\mu$m and 100\,$\mu$m, and less than 20\% at 160\,$\mu$m for PACS \citep{2010A&A...518L...2P}.  To obtain the flux of the blob, we used a target radius of $20 \arcsec$ on the difference maps.  The results are summarised in Table \ref{table : fluxes blob}.    In the difference maps, the resi\-duals from the disk subtraction become substantial, especially at 250\,$\mu$m.  The errors on the fluxes of the blob are hence calculated as the standard deviation on the flux in 20 circular apertures (with a radius of $20 \arcsec$) close to the subtracted disk.

To determine an upper limit for the flux of the blob in the PACS maps, we calculated the flux within a $10 \arcsec$ aperture at the position of the blob (in the original images) and subtracted the flux within a $10 \arcsec$ aperture on the opposite side of the star.  Scaling the difference in flux between the SW- and NE-aperture with the encircled energy function yields the upper limits listed in Table \ref{table : fluxes blob}.

\begin{table}
	\caption{Flux of the debris disk for PACS and SPIRE (calculated for a $60 \arcsec$ target radius).}
	\label{table : fluxes disk}
	\centering
	\begin{tabular}{cc}
		\hline \hline
		Wavelength [$\mu$m] & Flux [Jy] \\
		\hline 70 & 15.34 $\pm$ 0.8 \\
		100 & 9.34 $\pm$ 0.5 \\ 
		160 & 4.55 $\pm$ 0.5 \\
		250 & 1.88 $\pm$ 0.14 \\
		350 & 0.81 $\pm$ 0.06 \\
		500 & 0.37 $\pm$ 0.03 \\
		\hline
	\end{tabular}
\end{table}

\begin{table}
	\caption{Upper limits for the flux of the blob in the PACS images and the flux of the blob in the three SPIRE bands (calculated for a $20 \arcsec$ target radius), and the corresponding beam FWHM.}
	\label{table : fluxes blob}
	\centering
	\begin{tabular}{ccc}
		\hline \hline
		Wavelength [$\mu$m] & Flux [Jy] & Beam [$\arcsec$]\tablefootmark{a} \\
		\hline
		 70 & $< 0.39$         &  $5.26 \times 5.61$\\
		100 & $< 0.16$         &  $6.69 \times 6.89$\\
		160 & $< 0.15$         &  $10.46 \times 12.06$ \\
		250 & 0.081$\pm$ 0.027  &  18.2 \\ 
		350 & 0.050 $\pm$ 0.006 & 24.9 \\ 
		500 & 0.026 $\pm$ 0.004 & 36.3 \\ 
		\hline
	\end{tabular}
	\tablefoot{\tablefoottext{a}{Taken from the PACS Observer's Manual and SPIRE Observers' Manual (version 2.4).}}
\end{table}
 
To identify the background sources, we applied the HIPE \textit{sourceExtractorSussextractor} task to the original map at 250\,$\mu$m on the homogeneously covered region (with signal-to-noise threshold of 5.0). This task yields a list of sky coordinates and fluxes for the extracted sources, which can be fed into the task as priors when applying it to the two other original maps.  This forces the task to extract fluxes for the same sources in all three images.  Sources within a radius of $50 \arcsec$ from the star were removed from the list and 50 sources were left for analysis (see Fig.\,\ref{fig : background sources}).  Their sky positions and fluxes in the SPIRE bands are summarised in Table \ref{table : background galaxies}.

\section{Discussion}

\subsection{Comparison of the far-IR colours of the blob, the disk and the background sources}
\begin{figure}
	\resizebox{\hsize}{!}{\includegraphics{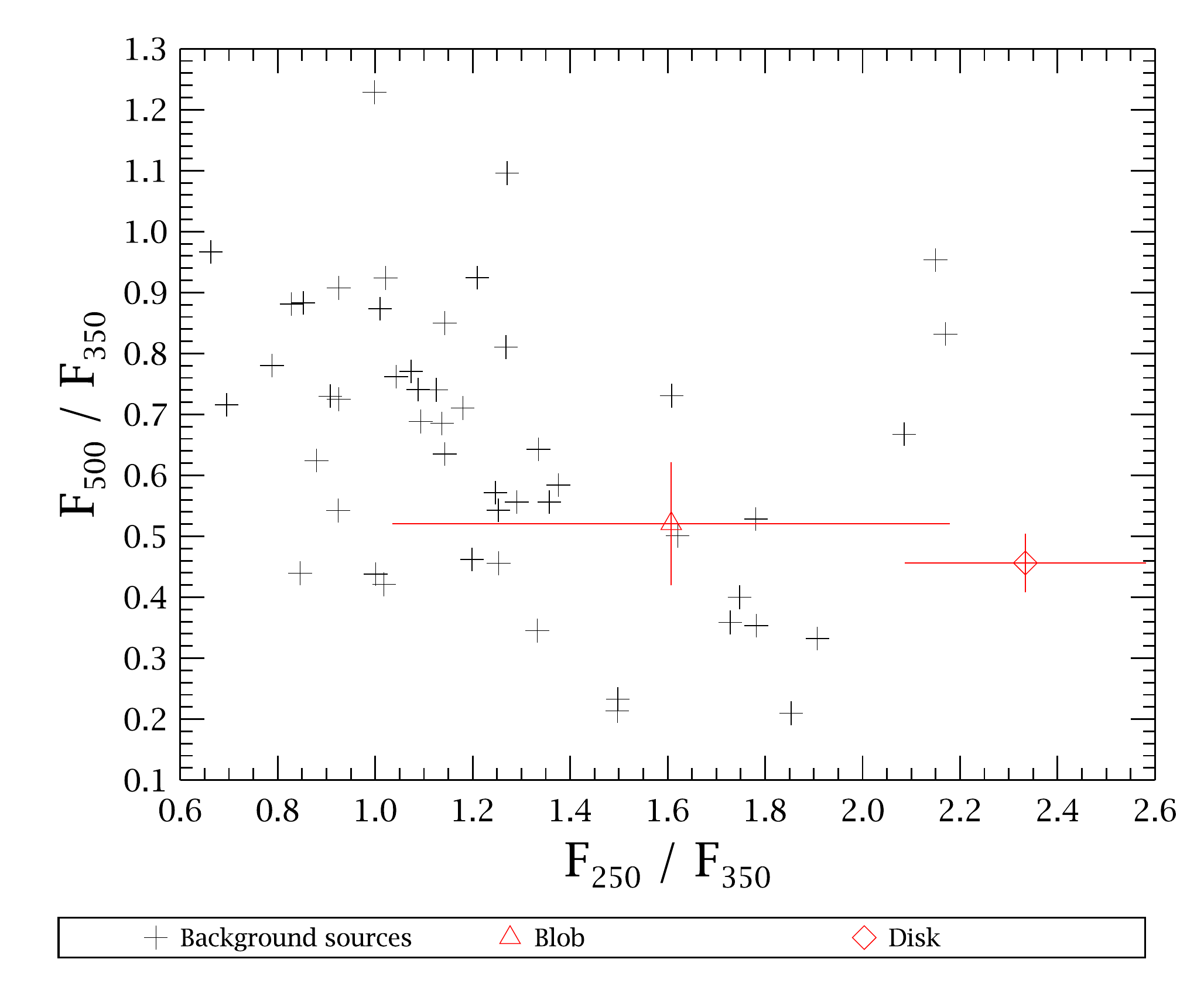}}
	\caption{Colour-colour diagram of the \object{$\beta$ Pic} debris disk, the blob and the background sources.  The error bars for the fluxes of the background galaxies are not visible on this scale.}
	\label{fig : fluxes SPIRE}
\end{figure}

Fig.\,\ref{fig : fluxes SPIRE} shows $F_{500}/F_{350}$ vs.\,$F_{250}/F_{350}$ colour-colour diagram for the debris disk, the blob and the background sources.  The colours of the blob are more similar to the colours of the background sources than the colours of the disk.  This is a first indication that the blob is a background source and not a structure in the disk, but no decisive conclusion can be drawn due to the relatively large error bars.

\subsection{Comparison of the size of the blob and the background sources}
A second argument that the blob is a background source was found in the comparison between the size of the blob and the size of the background sources.  The full-width at half-maximum (FWHM) of the background sources was determined in the three SPIRE bands by fitting a Gaussian to each of the sources in the original maps.  A Gaussian fit to the blob in the difference maps yielded the FWHM of the blob in the three bands.  The results are listed in Table\,\ref{table : SPIRE fwhm}.

The blob is not resolved, and its size is clearly consistent with the size of the background sources.  This is a second indication that the blob is a background source.

\begin{table}
	\caption{Median FWHM of the background sources and the FWHM of the blob in the three SPIRE bands.}
	\label{table : SPIRE fwhm}
	\centering
	\begin{tabular}{ccc}
		\hline \hline
		Wavelength [$\mu$m] & Background sources [$\arcsec$] & Blob [$\arcsec$] \\
		\hline
			250 & 18.7 $\pm$  5.5 & 23.4 \\
			350 & 25.7 $\pm$ 11.2 & 33.4 \\
			500 & 38.2 $\pm$ 19.5 & 37.2 \\
		\hline
	\end{tabular}
\end{table}

\subsection{Far-infrared SED of the blob and the disk, and interpretation as dust spectrum}
\begin{figure}
	\resizebox{\hsize}{!}{\includegraphics{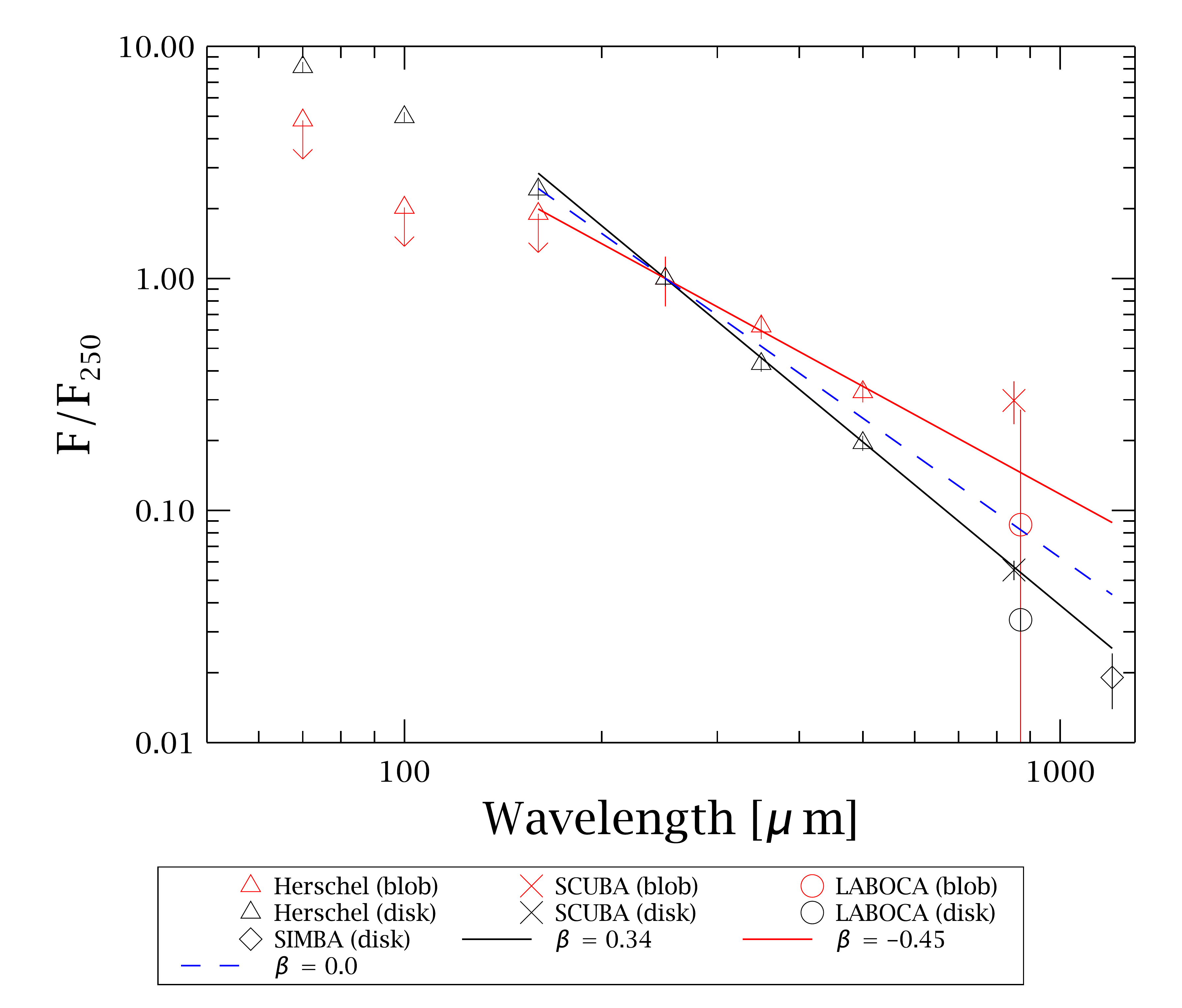}}
	\caption{Spectral energy distribution of the debris disk (in black) and the blob (in red), normalised to their repective flux at 250\,$\mu$m.  The \textit{Herschel} fluxes of the disk were obtained by integrating over a target radius of $60\arcsec$, the SPIRE fluxes of the blob by integrating over a target radius of $20 \arcsec$.  The PACS fluxes of the blob are upper limits, determined as described in Sect.\,\ref{section : image analysis}.  Additional flux points for the disk were taken from \citet{1998Natur.392..788H} (SCUBA: 850\,$\mu$m), \citet{2009A&A...508.1057N} (LABOCA: 870\,$\mu$m) and \citet{2003A&A...402..183L} (SIMBA: 1.2\,mm).  The SCUBA and SIMBA fluxes were integrated over a $40 \arcsec$ radius aperture while the LABOCA flux point was obtained by fitting a Gaussian to the disk. Additional flux points for the blob were taken from \citet{2000MNRAS.314..702D} (SCUBA : 850\,$\mu$m) and \citet{2009A&A...508.1057N} (LABOCA : 870\,$\mu$m).  Overplotted are three modified Rayleigh-Jeans laws ($F_{\nu} \propto \nu^{\beta + 2}$).  The spectral indices have been determined using the fluxes at wavelengths $\lambda \ge $160\,$\mu$m for the disk and $\lambda \ge $250\,$\mu$m for the blob, with the inverse of the errors on the fluxes used as weight.  Note the large error bar for the 870\,$\mu$m flux point of the blob.  The $\beta = 0$ curve is shown for comparison.}
	\label{fig : SED}
\end{figure}

Another test for the hypothesis  that the blob is a background source are the relative spectral energy distributions (SEDs) of the disk and the blob.  Fig.\,\ref{fig : SED} shows the SED of the debris disk with fluxes obtained with \textit{Herschel}, SCUBA, LABOCA, and SIMBA, normalised to the flux of the disk at 250\,$\mu$m, as well as the flux points for the blob, normalised to the flux of the blob at 250\,$\mu$m (in red).  Overplotted are three modified Rayleigh-Jeans laws ($F_{\nu} \propto \nu^{\beta + 2}$).  The spectral index $\beta$ describes the mean dust opacity ($\kappa \propto \nu^{\beta}$) and is an indicator of the grain size distribution of the dust.  The SED of the disk is best described by a spectral index $\beta = 0.34$ (implying (sub)$\mu$m grains), while a negative spectral index is needed to fit the SED of the blob ($\beta = -0.45$ produces the best result).  The latter can only be explained by a steep temperature gradient that cannot occur on the scale the blob would have if it were a structure in the disk (i.e., about 470\,AU in dia\-meter).

The submillimetre fluxes of the blob and the disk, as detected by SCUBA, LABOCA, and SIMBA, are summarised in Table \ref{table : submm fluxes}.

\subsection{Determination of the redshift}
\begin{figure}
	\resizebox{\hsize}{!}{\includegraphics{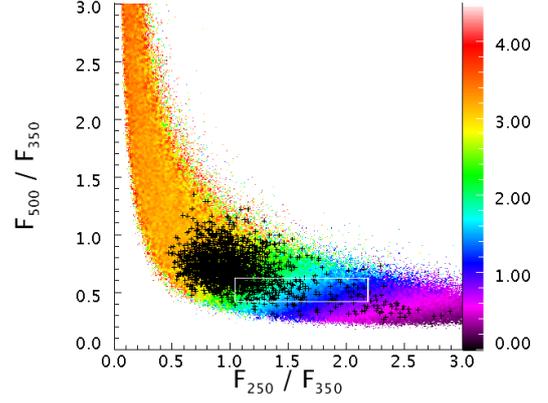}}
	\caption{Colour-colour diagram of the \textit{Herschel}-ATLAS sources \citep{2010A&A...518L...9A}.  The coloured background indicates the average redshift for $10^6$ randomly generated model SEDs.  The SPIRE colours of background sources can be overplotted to determine their redshift.  Other object types (e.g.\,the debris disk itself) should not be overplotted.  The small crosses are the 1947 galaxies detected in the context of the ATLAS open-time key programme (with a significance greater than $5 \sigma$ at 350\,$\mu$m and $3 \sigma$ at 250 and 500\,$\mu$m).  The overplotted rectangle indicates the position of the blob in this colour-colour space. The redshift in this rectangle lies between 0.78 and 2.6.}
	\label{fig : redshift}
\end{figure}

Assuming the blob really is a background source, we can estimate its redshift based on the far-infrared fluxes. \citet{2010A&A...518L...9A} investigated the redshift distribution of submillimetre galaxies in the \textit{Herschel}-ATLAS survey. Their $F_{500}/F_{350}$ vs.\,$F_{250}/F_{350}$ diagram shows the average redshift in that colour-colour space.  In Fig.\,\ref{fig : redshift}, we have overplotted the data\-point for the blob.  Within the error bars, we find redshifts between 0.78 and 2.55.

Assuming the blob is a background galaxy with redshift $z = 1.6$, we derive a luminosity of the order of  $10^{26} \textrm{W Hz}^{-1}$ at 250\,$\mu$m (from the integrated flux at that wavelength). The luminosity distance $d_L$ we used for this calculation was derived from the relationship in Fig.\,1 of \citet{2005A&A...429..807C}.  The luminosity function of \citet[][their Fig.\,4]{2010A&A...518L..23E} shows this is an unlikely high value.  Higher redshifts would result in even higher luminosities.  Because this is improbable, we estimate $z$ to be no higher than 1.6.

\begin{figure}
	\resizebox{\hsize}{!}{\includegraphics{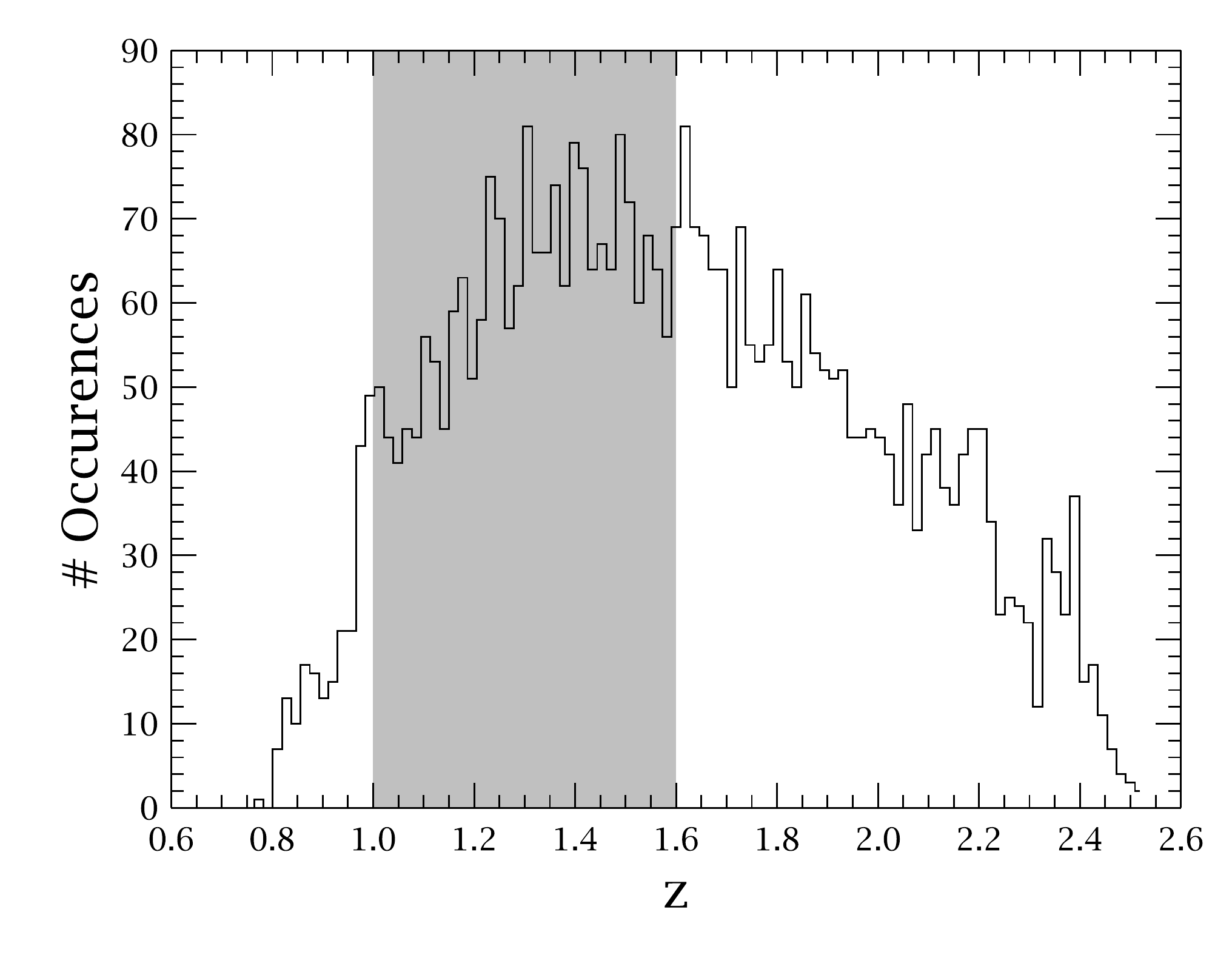}}
	\caption{Histogram of the redshift distribution in the rectangle marked in Fig.\,\ref{fig : redshift}.  Values range between 0.78 and 2.55.  The shaded region indicates the redshift to which we have narrowed down the region.}
	\label{fig : histogram redshift}
\end{figure}

Rejecting the 5\% lowest values for the redshift within the overplotted box (see Fig.\,\ref{fig : histogram redshift} for the histogram), we estimate $z$ to be at least 1.0.

\section{Conclusions}
We have investigated the nature of the blob that has been detected to the southwest of the debris disk of $\object{\beta$ Pic} not only with \textit{Herschel}, but also with SCUBA, LABOCA, and SIMBA.  By comparing the submillimetre colours of the blob with the colours of the debris disk itself and with the background sources that are present in the images, we conclude that the blob is not a structure in the disk, as previously suggested, but  a background galaxy.  Support for this hypothesis was found from differences in the spectral indices and SEDs between the disk and the blob, and the fact that the size of the blob is consistent with the size of the background sources.  Comparing the colour of the blob to the colour-colour diagram from \citet{2010A&A...518L...9A} yielded a rough estimate for the redshift of the blob ($z$ between 0.78 and 2.55), which we narrowed down to $z$ between 1.0 and 1.6.

Ultimately, the nature of the blob could be probed by spectroscopy.  The redshift of a galaxy can be determined based on the CO emission, because the spacing between two successive transitions shrinks with redshift (as $z+1$).  The large total bandwidth and high angular resolution make ALMA the ideal instrument to study the CO emission of the blob, and determine its redshift.  For a normal galaxy like the Milky Way, with $z=3$, the study of the CO emission takes less than 24\,h observing time \citep{ALMA}.

Band 3, the most sensitive available band, will show CO\,2-1 transitions for $z = 1-2$.  Higher (lower) redshifts would move CO\,2-1 out of the band, and CO\,3-2 (CO\,1-0) into the bandpass.  Observing additional lines in, e.g., band 7 (with the penalty of lower sensitivity), or JVLA (for CO\,1-0) can rule out any ambiguity.

\begin{acknowledgements}
The authors thank the referee for constructive comments that have improved the paper in several aspects.
This work was supported in part by the Belgian Federal Science Policy Office via the PRODEX Programme of ESA.
PACS has been developed by a consortium of institutes led by MPE (Germany) and including UVIE (Austria);
KU Leuven, CSL, IMEC (Belgium); CEA, OAMP (France); MPIA (Germany); IFSI, OAP/AOT, OAA/CAISMI, LENS, SISSA (Italy); IAC (Spain). This development has been supported by the funding agencies BMVIT (Austria), ESA PRODEX (Belgium), CEA/CNES (France), DLR (Germany), ASI (Italy), and CICT/MCT (Spain).
The authors thank A.~Amblard for providing the data to generate Fig.\ref{fig : redshift}.
\end{acknowledgements}

\bibliographystyle{aa}
\bibliography{biblioBetaPic}


\section*{Photometry of the background galaxies in the SPIRE wavelength bands}
Table \ref{table : background galaxies} summarises the positions and photometry in the three SPIRE bands for the extracted background galaxies.

\begin{table*}
	\caption{Positions and photometry of the background sources in the SPIRE wavelength bands.}
	\label{table : background galaxies}
	\centering
	\begin{tabular}{rrrrrrr}
		\hline \hline index & $\alpha$ [h m s] & $\delta \ [\degr \ \arcmin \ \arcsec] $  & $F_{250}$ [mJy] & $F_{350}$ [mJy] & $F_{500}$ [mJy] \\
		\hline
 1  & 5 \  47 \  02.97  &  -51 \  03 \  46.27  &  86.55  &  53.41  &  26.75  \\
 2  & 5 \  47 \  14.96  &  -51 \  08 \  11.93  &  88.56  &  84.93  &  64.70  \\
 3  & 5 \  47 \  08.37  &  -51 \  05 \  21.58  &  66.35  &  61.80  &  47.62  \\
 4  & 5 \  47 \  41.45  &  -51 \  08 \  31.18  &  49.83  &  27.97  &  09.89  \\
 5  & 5 \  47 \  10.16  &  -51 \  07 \  52.80  &  51.50  &  47.34  &  35.06  \\
 6  & 5 \  47 \  29.32  &  -51 \  07 \  03.24  &  44.35  &  39.41  &  29.18  \\
 7  & 5 \  47 \  19.58  &  -51 \  01 \  25.95  &  44.32  &  47.92  &  34.75  \\
 8  & 5 \  47 \  03.42  &  -51 \  06 \  00.39  &  40.79  &  30.55  &  19.63  \\
 9  & 5 \  47 \  10.92  &  -51 \  03 \  52.90  &  36.46  &  39.41  &  35.76  \\
10  & 5 \  46 \  46.97  &  -51 \  08 \  54.68  &  56.67  &  42.52  &  14.68  \\
11  & 5 \  47 \  05.71  &  -51 \  05 \  47.38  &  36.85  &  30.46  &  28.16  \\
12  & 5 \  47 \  01.50  &  -51 \  03 \  11.85  &  36.67  &  40.39  &  29.48  \\
13  & 5 \  47 \  24.11  &  -51 \  03 \  07.39  &  38.46  &  25.69  &  05.48  \\
14  & 5 \  47 \  11.78  &  -51 \  04 \  42.97  &  31.45  &  31.49  &  38.70  \\
15  & 5 \  47 \  08.21  &  -51 \  04 \  07.02  &  32.21  &  25.35  &  27.79  \\
16  & 5 \  47 \  00.32  &  -50 \  59 \  00.80  &  32.99  &  29.02  &  19.88  \\
17  & 5 \  47 \  06.40  &  -51 \  00 \  46.39  &  34.05  &  24.75  &  14.45  \\
18  & 5 \  47 \  05.10  &  -51 \  04 \  24.99  &  31.98  &  29.24  &  20.13  \\
19  & 5 \  46 \  51.74  &  -51 \  01 \  18.91  &  32.19  &  32.15  &  14.07  \\
20  & 5 \  47 \  24.20  &  -51 \  01 \  12.17  &  31.97  &  18.50  &  06.63  \\
21  & 5 \  47 \  26.50  &  -51 \  08 \  47.06  &  34.41  &  16.50  &  11.01  \\
22  & 5 \  46 \  53.13  &  -51 \  02 \  39.18  &  27.32  &  14.74  &  03.09  \\
23  & 5 \  47 \  46.42  &  -51 \  03 \  22.98  &  28.83  &  23.02  &  12.49  \\
24  & 5 \  47 \  44.64  &  -50 \  59 \  59.36  &  29.00  &  15.20  &  05.05  \\
25  & 5 \  46 \  58.97  &  -51 \  03 \  37.51  &  25.57  &  22.38  &  19.02  \\
26  & 5 \  46 \  55.88  &  -50 \  59 \  19.69  &  26.61  &  26.05  &  24.07  \\
27  & 5 \  47 \  28.34  &  -51 \  00 \  31.02  &  24.14  &  27.45  &  17.13  \\
28  & 5 \  47 \  11.41  &  -51 \  07 \  22.61  &  26.36  &  31.83  &  28.04  \\
29  & 5 \  47 \  24.42  &  -51 \  08 \  53.03  &  23.56  &  18.58  &  15.06  \\
30  & 5 \  47 \  03.81  &  -51 \  01 \  38.75  &  23.51  &  18.86  &  10.78  \\
31  & 5 \  47 \  01.02  &  -51 \  00 \  24.24  &  22.22  &  12.47  &  06.59  \\
32  & 5 \  47 \  28.00  &  -51 \  07 \  44.27  &  21.76  &  19.04  &  12.09  \\
33  & 5 \  46 \  54.81  &  -51 \  04 \  45.89  &  20.21  &  29.07  &  20.81  \\
34  & 5 \  47 \  46.63  &  -51 \  02 \  50.88  &  18.67  &  15.82  &  11.24  \\
35  & 5 \  47 \  28.09  &  -51 \  01 \  24.72  &  19.02  &  15.18  &  06.92  \\
36  & 5 \  47 \  48.58  &  -51 \  04 \  52.28  &  19.67  &  12.23  &  08.94  \\
37  & 5 \  47 \  10.79  &  -51 \  02 \  04.77  &  16.18  &  13.50  &  06.24  \\
38  & 5 \  47 \  03.83  &  -51 \  00 \  16.54  &  18.47  &  18.14  &  07.64  \\
39  & 5 \  47 \  45.58  &  -51 \  04 \  52.72  &  16.10  &  07.42  &  06.17  \\
40  & 5 \  47 \  22.00  &  -51 \  05 \  27.15  &  13.79  &  09.21  &  02.14  \\
41  & 5 \  47 \  22.66  &  -51 \  07 \  32.10  &  14.72  &  08.42  &  03.36  \\
42  & 5 \  47 \  09.78  &  -51 \  00 \  11.45  &  12.65  &  09.32  &  05.18  \\
43  & 5 \  47 \  13.95  &  -50 \  58 \  56.53  &  13.08  &  06.08  &  05.80  \\
44  & 5 \  46 \  50.31  &  -50 \  59 \  59.64  &  14.08  &  17.86  &  13.94  \\
45  & 5 \  47 \  30.64  &  -51 \  08 \  37.85  &  10.32  &  12.11  &  10.70  \\
46  & 5 \  47 \  03.08  &  -51 \  05 \  19.24  &  10.16  &  10.06  &  08.78  \\
47  & 5 \  46 \  52.54  &  -50 \  59 \  56.15  &  11.36  &  17.14  &  16.57  \\
48  & 5 \  46 \  57.46  &  -51 \  08 \  35.38  &  10.75  &  12.72  &  05.59  \\
49  & 5 \  47 \  07.62  &  -51 \  01 \  52.54  &  10.11  &  07.83  &  04.35  \\
50  & 5 \  46 \  55.04  &  -51 \  08 \  30.49  &  10.55  &  11.42  &  06.19  \\
	\hline
	\end{tabular}
\end{table*}

\section*{Submillimetre fluxes of the disk and blob}
In Table \ref{table : submm fluxes} the submillimetre fluxes of the disk and the blob as measured with SCUBA, LABOCA, and SIMBA are listed.

\begin{table}
	\caption{Submillimetre fluxes of the disk and the blob as detected by SCUBA, LABOCA, and SIMBA. The blob was not observed at 1.2\.mm}
	\label{table : submm fluxes}
	\centering
	\begin{tabular}{cccc}
		\hline \hline
		wavelength [$\mu$m] & flux disk [mJy] & flux blob [mJy] & References \\
		\hline 850 & 104.3 $\pm$ 10.0 &  24 $\pm$ 5  & $1, 2$ \\
		870 & 63.6 $\pm$ 6.7 & 7.0 $\pm$ 14.9 & 3 \\ 
		1200 & 35.9 $\pm$ 9.7 & / &  4 \\ 
		\hline
	\end{tabular}
	\tablebib{(1)~\citet{1998Natur.392..788H}; (2)~\citet{2000MNRAS.314..702D}; (3)~\citet{2009A&A...508.1057N}; (4)~\citet{2003A&A...402..183L}
}
\end{table}
\end{document}